\documentclass{jfm}
\usepackage{graphicx}
\usepackage{epstopdf, epsfig}
\usepackage{natbib}
\usepackage{color}
\usepackage[dvipsnames]{xcolor}
\usepackage{amssymb}
\usepackage{amsmath}
\usepackage{bm} 
\usepackage{latexsym}

\newcommand\Reytau{\mathrm{Re}_{\tau}}
\newcommand\dd{\mathrm{d}}

\shorttitle{Superpipe data reanalyzed}
\shortauthor{P. A. Monkewitz}

\title{Derivation of Pitot corrections for the Zagarola \& Smits Superpipe data\\and their composite fit}

\author{Peter A. Monkewitz
  \corresp{\email{peter.monkewitz@epfl.ch}}}

\affiliation{Ecole Polytechnique F\'ed\'erale de Lausanne (EPFL), CH-1015, Lausanne, Switzerland}

\begin{document}

\maketitle   

\begin{abstract}
Rejected by JFM Rapids, May 2018
\\*
\\*
The original turbulent pipe flow experiments in the Princeton ``Superpipe'' by \cite{ZS97, ZS98} at unprecedented laboratory Reynolds numbers have started an ongoing vigorous debate on the logarithmic law in the mean velocity profile $U^+(y^+)$ and the intimately related question of Pitot probe corrections for mean shear, viscous effects and turbulence level. Considering that the Pitot probe diameter $d^+$ exceeded 7000 wall units at the highest Reynolds number, the various \textcolor{black}{traditional Pitot} corrections had to be extended into uncharted territory \textcolor{black}{where they may no longer be additive}. In this note, the inverse approach is adopted, where the net result of all the corrections is assumed to be compatible with the model for $U^+$ developed by \cite{Monk17}. The latter has an inner part which is, up to higher order corrections, identical to the zero pressure gradient turbulent boundary layer profile and switches around $y^+_{\mathrm{break}} \approx 400$ to a logarithmic overlap layer with a K\'arm\'an ``parameter'' $\kappa$ \textcolor{black}{that depends on pressure gradient and possibly on other flow parameters}. The simplicity of the resulting global Pitot correction proportional to $(d^+)^{0.9}(R^+)^{-0.4}$, with only two fitting parameters, indirectly supports this model. \textcolor{black}{Based on the required equality of the overlap and centerline $\kappa$'s, it is furthermore shown that $y^+_{\mathrm{break}}$ must be a constant.}
\textcolor{black}{Finally,} the outer ``wake'' part of the profile is argued to be asymptotically linear between the wall and about half the pipe radius. This gives rise to a linear higher order tail $\propto y^+/R^+$ in the logarithmic overlap layer, which has been \textcolor{black}{the subject} of asymptotic analysis over the last decades.
\end{abstract}

\section{\label{sec:intro}The problem with the K\'arm\'an constant in pipe flow}

In early pipe flow experiments, the emphasis has been on the scaling of the centerline velocity and the friction factor with Reynolds number. As measurement techniques have evolved, attention has shifted towards the ``law of the wall'' $U^+ = (1/\kappa)\ln(y^+) + B$, where $\kappa$ is the celebrated K\'arm\'an ``constant''. In the last years the value of $\kappa$ in pipes has closely approached the ``most popular'' value of 0.384 for the zero-pressure-gradient boundary layer \citep{Furuichi15,CiclopeRS}, which seemingly supports the claim of \textcolor{black}{\citep[see e.g.][and others]{MMHS13, KPM17}}, that $\kappa = 0.39$ is universal for zero-pressure-gradient boundary layers, \textcolor{black}{ pipe and channel flows}.

Here and in the following, ``+'' superscripts indicate non-dimensionalization with wall units $\widehat{u}_\tau \equiv(\widehat{\tau}_{\mathrm{wall}}/\widehat{\rho})^{1/2}$ and \textcolor{black}{$\widehat{\ell} \equiv \widehat{\nu}/\widehat{u}_\tau$}, where hats indicate dimensional quantities. The relevant Reynolds number for pipe flow is $\Reytau \equiv R^+ = (\widehat{R}\,\widehat{u}_\tau/\widehat{\nu})$, with $\widehat{R}$ the pipe radius.

\begin{figure}
\center
\includegraphics[width=0.65\textwidth]{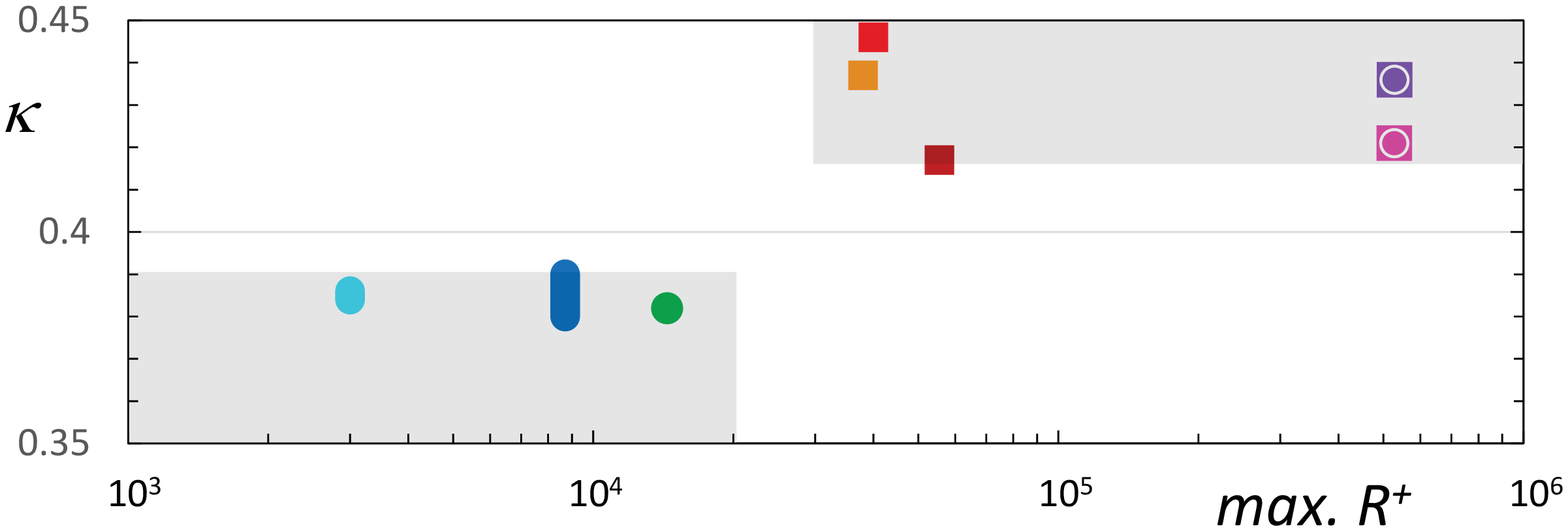}
\caption{(color online) $\kappa$'s determined by different authors from near-wall $U^+(y^+)$ profiles ($\bullet$) and from the centerline velocity $U^+_{\mathrm{CL}}(R^+)$ ($\blacksquare$) versus maximum $R^+$ of the respective experiment: \textcolor{cyan}{$\bullet$}, \cite{Monty_thesis}; \textcolor{blue}{$\bullet$}, \cite{zanoun2007}; \textcolor{green}{$\bullet$}, \cite{Furuichi15}; \textcolor{orange}{$\blacksquare$}, \cite{FioriniPhD}; \textcolor{red}{$\blacksquare$}, \cite{TSFP17}; \textcolor{brown}{$\blacksquare$}, \cite{Nikuradse-pipe}; \textcolor{violet}{$\bullet\, \blacksquare$}, \cite{ZS98} and \textcolor{purple}{$\bullet\, \blacksquare$}, \cite{MLJMS04}. The gray boxes emphasize the separation of values obtained from low and high Reynolds number experiments.}
\label{Fig:kappas}
\end{figure}

Since the pioneering work of \cite{Coles56} it has been recognized that in turbulent wall-bounded flows, the logarithmic overlap layer or common part $U^+_{\mathrm{cp}}(y^+) = \ln(y^+)/\kappa + B$ of the ``inner'' and ``outer'' asymptotic expansions of the mean velocity necessarily entails a free stream velocity $U^+_{\infty}(\Reytau)$ or centerline velocity $U^+_{\mathrm{CL}}(\Reytau)$ of the form $\ln(\Reytau)/\kappa + C$, with the \textit{same} logarithmic slope $(1/\kappa)$. While in the zero-pressure-gradient turbulent boundary layer, henceforth abbreviated ZPG TBL, the equality of $\kappa$'s obtained from the overlap layer and from the evolution of the free-stream velocity $U^+_\infty$ with Reynolds number has remained non-controversial \citep[see e.g.][]{MCN07, Metal2010}, the $\kappa$'s extracted by different authors from pipe flow profiles have not converged to a generally agreed value, as already discussed by \cite{NagibChauhan2008} and shown in figure \ref{Fig:kappas}. What is striking in this figure, is the rather clear separation of reported $\kappa$ values according to the Reynolds number range of the experiment: while the $\kappa$'s remain within 0.005 of the widely accepted ZPG TBL value of 0.384 for $R^+_{\mathrm{max}} \lesssim 2\times 10^4$, they jump to 0.42 and beyond in experiments with $R^+_{\mathrm{max}} \gtrsim 2\times 10^4$.

Apart from the data of \cite{Nikuradse-pipe}, difficult to assess, the Princeton Superpipe data were for a long time the only laboratory data beyond an $R^+$ of $2\times 10^4$ and the $\kappa$'s from the log-law and from the centerline were in agreement, as they should be. However, the log-laws identified by \cite{ZS98} with $\kappa = 0.436$ and by \cite{MLJMS04} with $\kappa = 0.421$ only started beyond a $y^+$ of the order of \textcolor{black}{$500$}, as opposed to $150-200$ in pipe experiments at lower Reynolds numbers and in the ZPG TBL. This gave rise to extended controversies about Pitot corrections \citep[see e.g.][]{perry2001,Pitot13,VinDN16} and to speculations about a ``mixing transition'' of pipe flow at $R^+ = \mathcal{O}(10^4)$ \citep{MZS05}. Only recently, \cite{Monk17} proposed a resolution of this conundrum. He showed that in pipe and channel flows, where the effect of pressure gradient on the near-wall momentum balance is weak, mean velocity profiles are well described by the ZPG TBL profile up to $y^+ \approx 400-500$, implying that the pipe profile includes the beginning of the ZPG TBL log-law with $\kappa_0 = 0.384$. Beyond this wall distance, the inner pipe profile veers off to the \textcolor{black}{``true''} overlap log-law with a pipe-specific $\kappa$ significantly higher than 0.384. \textcolor{black}{Considering that beyond $Y \approx 0.05$, with $Y \equiv y^+/R^+$ the outer coordinate, the overlap log-law becomes progressively contaminated by the wake, the clean overlap log-law only becomes visible for $400 \ll 0.05 R^+$, which corresponds in practice to $R^+ \gtrsim 2\times 10^4$, as seen in figure \ref{Fig:kappas}.}

\section{\label{sec:corrs}The Pitot and other corrections for the Zagarola \& Smits data}

\subsection{\label{sec:Hama}Roughness correction}

At the higher $R^+$, the Hama-like roughness correction of \cite{Monk17}
\begin{equation}
\Delta U^{+}_{\mathrm{rough}} = \frac{2}{\kappa}\,\ln[1 + (0.14\,k_s^+)^2] \quad \mbox{with} \quad \widehat{k}_s = 0.45\,\mu\mathrm{m} \quad ,
\label{Hama}
\end{equation}
has been applied, which is in line with the investigation of \citet{Allen05} \textcolor{black}{and significantly affects the profiles only for $R^+ \gtrapprox 2\times 10^5$}.

\subsection{\label{sec:corrPitot}``Reverse engineering'' of the Pitot correction for the Zagarola \& Smits data}

The resolution of the discrepancy between $\kappa$'s from low and high Reynolds number experiments by \cite{Monk17} is supported by the success of using his idea ``backwards'', i.e. to assume that the near-wall profiles are identical in pipe and ZPG TBL, and modelled by the Musker profile modified by \cite{Chauhan09}:
\begin{align}
\label{Musker}
U^{+\,\mathrm{(ZPG)}}_{\mathrm{inner}} \cong \frac{1}{\kappa_0}\ln\left(\frac{y^+ + a}{a}\right) - \frac{\gamma^2}{a(4\alpha + a)} \left\{ (4\alpha - a)\,\ln\left(\frac{a \sqrt{(y^+ - \alpha)^2 + \beta^2}}{\gamma (y^+ + a)}\right) \right.\nonumber \\
+ \left.\frac{\alpha (4\alpha - 5a)}{\beta} \textcolor{black}{\left[\arctan \left(\frac{y^+ - \alpha}{\beta}\right) + \arctan\left(\frac{\alpha}{\beta}\right)\right]}\right\} + \frac{\exp\left[- \ln^2(y^+/30)\right]}{2.85}
\end{align}
where $\alpha = (a - 1/\kappa_0)/2$, $\beta = \sqrt{2a\alpha - \alpha^2}$ and $\gamma = \sqrt{\alpha^2 + \beta^2}$.
With $\kappa_0 = 0.384$ and the parameter $a = 10.35$, the Musker profile (\ref{Musker}) asymptotes to $U^+ \sim (1/0.384)\,\ln(y^+) + 4.21$, as in \cite{Monk17} \textcolor{black}{(Note that in appendix A of \cite{Monk17} the above two arctan have been combined into one, which requires a branch switch at $y^+ = 20.7$, i.e. the addition of $\pi$ for $y^+ \geq 20.7$)}.

\begin{figure}
\center
\includegraphics[width=0.65\textwidth]{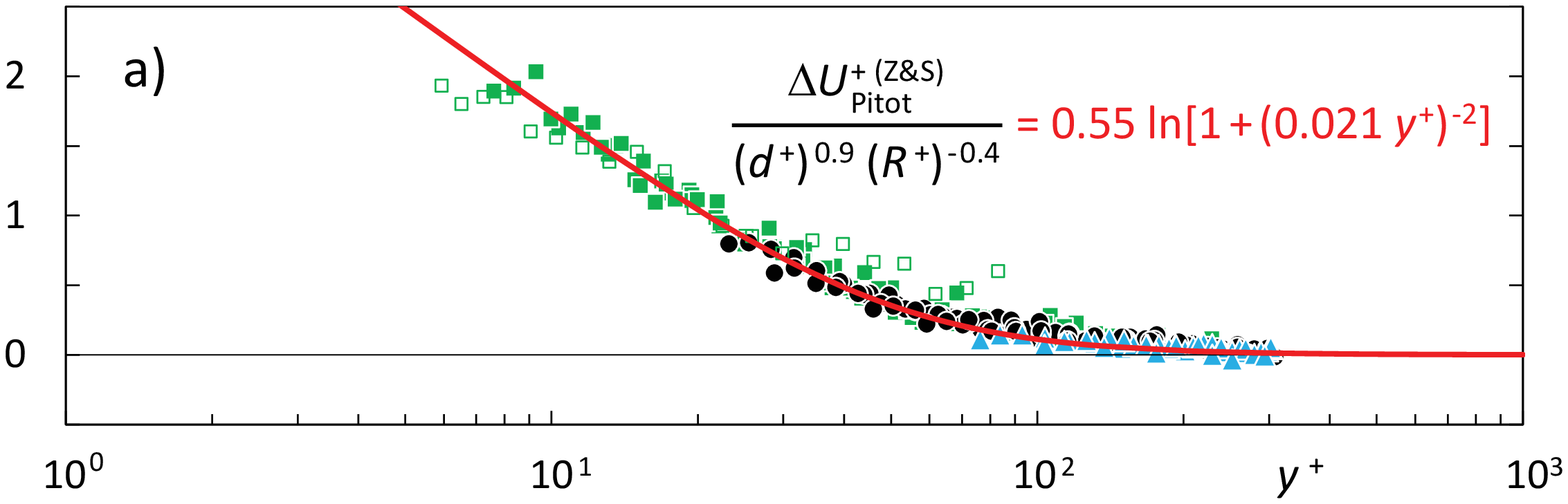}
\includegraphics[width=0.45\textwidth]{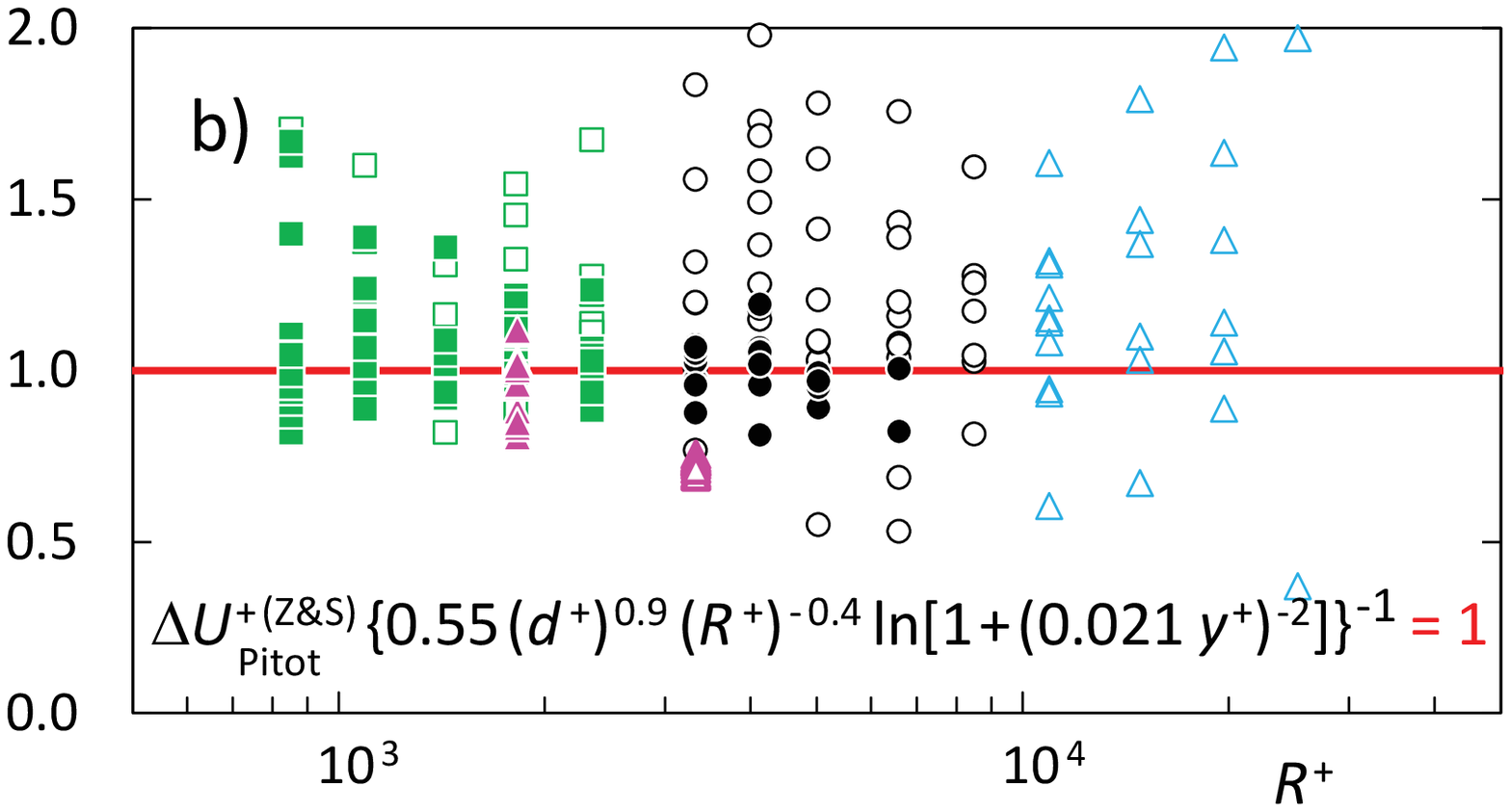}
\includegraphics[width=0.45\textwidth]{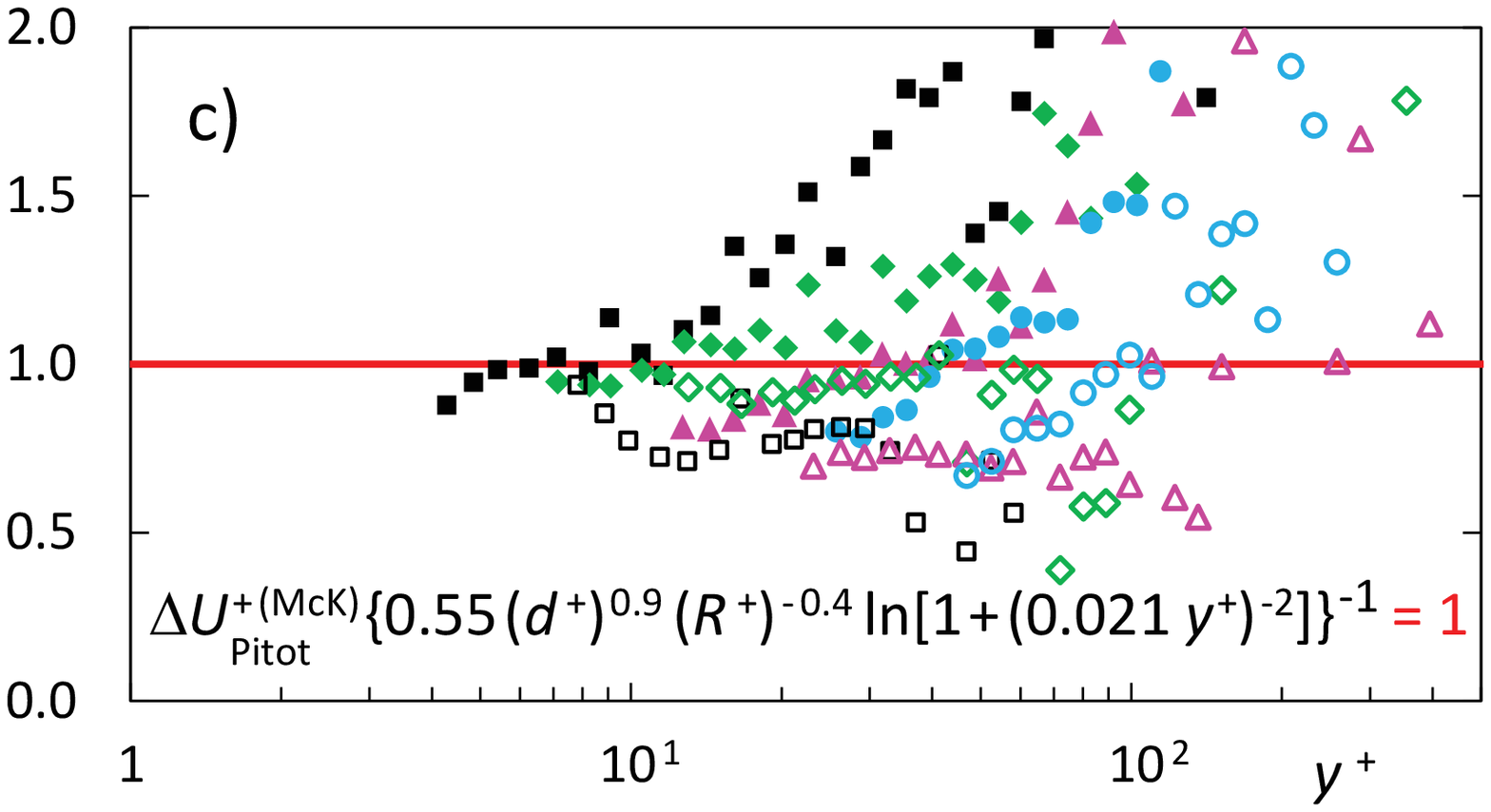}
\caption{(color online)\ (a) Difference $\Delta U^{+\,\mathrm{(Z\&S)}}_{\mathrm{Pitot}}$ (equ. \ref{Zagcorr}) between the uncorrected pipe velocities $U^{+\,\mathrm{(Z\&S)}}_{\mathrm{uncorr}}$ of \cite{ZS98} and the Musker-Chauhan fit $U^{+\,\mathrm{(ZPG)}}_{\mathrm{inner}}$ for the ZPG TBL inner expansion (equ. \ref{Musker}), scaled by $(d^+)^{0.9}(R^+)^{-0.4} \equiv 0.0213 (R^+)^{0.5}$. Data range $0 \leq y^+ \leq \mathrm{min}[300, 0.1\,R^+]$ containing data up to $R^+\approx 4\times 10^4$. Data symbols as in fig. \ref{Fig:Plog1} except for the lowest $R^+=851$ identified by \textcolor{green}{$\Box$}. \textcolor{red}{\textbf{---}}, fit by equation (\ref{Zagcorr}).
\newline (b) Scaled difference $\Delta U^{+\,\mathrm{(Z\&S)}}_{\mathrm{Pitot}}$ versus $R^+$ for the Z\&S data with $d/R = 0.0139$. Solid and open symbols correspond to $y^+ \leq 50$ and $50 < y^+ \leq 300$, respectively. The McKeon data of panel (c) for the same $d/R = 0.0139$ are included as purple triangles.
\newline (c) Analogous scaled difference $\Delta U^{+\,\mathrm{(McK)}}_{\mathrm{Pitot}}$ between the data from Appendix C of \cite{McK_thesis} and $U^{+\,\mathrm{(ZPG)}}_{\mathrm{inner}}$ versus $y^+$, for $R^+ = 1825$ and Pitot O.D.'s of 0.3mm ($\blacksquare$), 0.5mm (\textcolor{green}{$\blacklozenge$}), 0.9mm (\textcolor{purple}{$\blacktriangle$}) and 1.8mm (\textcolor{blue}{$\bullet$}); Corresponding open symbols, data for $R^+ = 3328$.}
\label{Fig:corr}
\end{figure}

The new global Pitot correction for the raw Superpipe data of \cite{ZS98} (available at https://smits.princeton.edu/zagarola/ and henceforth referred to as Z\&S data) is now obtained by fitting the difference $\Delta U^{+\,\mathrm{(Z\&S)}}_{\mathrm{Pitot}} \equiv [U^{+\,\mathrm{(Z\&S)}}_{\mathrm{uncorr}} - U^{+\,\mathrm{(ZPG)}}_{\mathrm{inner}}]$ between the uncorrected Z\&S data and the inner expansion in the ZPG TBL, modeled by equation (\ref{Musker}). This difference $\Delta U^{+\,\mathrm{(Z\&S)}}_{\mathrm{Pitot}}$ is found to \textcolor{black}{depend logarithmically on $y^+$ and scale as $(d^+)^{m}(R^+)^{(0.5 - m)} = (0.0139)^m (R^+)^{0.5}$ for the fixed ratio $d/R = 0.0139$ ($\widehat{d} = 0.9$mm) used by \cite{ZS98}. As seen in figure \ref{Fig:corr}a, the resulting collapse of the Z\&S data is exceptionally good and nicely fitted by the simple global Pitot correction}
\textcolor{black}{\begin{equation}
\Delta U^{+}_{\mathrm{Pitot}} \equiv U^{+\,\mathrm{(P)}}_{\mathrm{uncorr}} - U^{+\,\mathrm{(ZPG)}}_{\mathrm{inner}} = 0.55 \,(d^+)^{0.9}(R^+)^{-0.4} \ln[1 + (0.021 y^+)^{-2}]
\label{Zagcorr}
\end{equation}}
\textcolor{black}{The scaling of $\Delta U^{+\,\mathrm{(Z\&S)}}_{\mathrm{Pitot}}$ with $(R^+)^{0.5}$ is further verified in figure \ref{Fig:corr}b for the data below $y^+$ of 50, where the normalizing factor (the RHS of equ. \ref{Zagcorr}) is sufficiently larger than the combined uncertainties of the data and of the Musker-Chauhan fit (\ref{Musker}). Finally, the exponent $m \approxeq 0.9$ in $(d^+)^{m}(R^+)^{(0.5 - m)}$ is found with the help of data in Appendix C of \cite{McK_thesis}, obtained with four different Pitot diameters at two $R^+$ of 1825 and 3328, and only corrected for static pressure errors. The normalized $\Delta U^{+\,\mathrm{(McK)}}_{\mathrm{Pitot}}$ resulting from the choice $m=0.9$ are seen in figure \ref{Fig:corr}c to scatter considerably for $y^+ \gtrapprox 50$ (even earlier for the smallest Pitot diameter of {0.3\,}mm), but $m$ will not be needed in the following which deals exclusively with the Z\&S Superpipe data.}

\textcolor{black}{In conclusion, since} equation (\ref{Zagcorr}) models the combined effect of different physical phenomena, no simple explanation for the proportionality of $\Delta U^{+}_{\mathrm{Pitot}}$ to $(d^+)^{0.9}(R^+)^{-0.4}$ nor its logarithmic dependence on $y^+$ can be offered. One can only speculate about the reasons for the \textcolor{black}{superior data collapse obtained by the present} ``black box'' approach, \textcolor{black}{as compared to} the individual, physics-based corrections of \cite{ZS98}, \cite{mckeonpitot03}, \cite{Pitot13} and \cite{VinDN16}. One possibility is that individual corrections may no longer be additive for very large $d^+$, which exceed 7000 at the highest $R^+$.

\section{\label{sec:fit}Composite expansion for the mean velocity}

\subsection{\label{sec:inner}The inner profile}

\begin{figure}
\center
\includegraphics[width=0.8\textwidth]{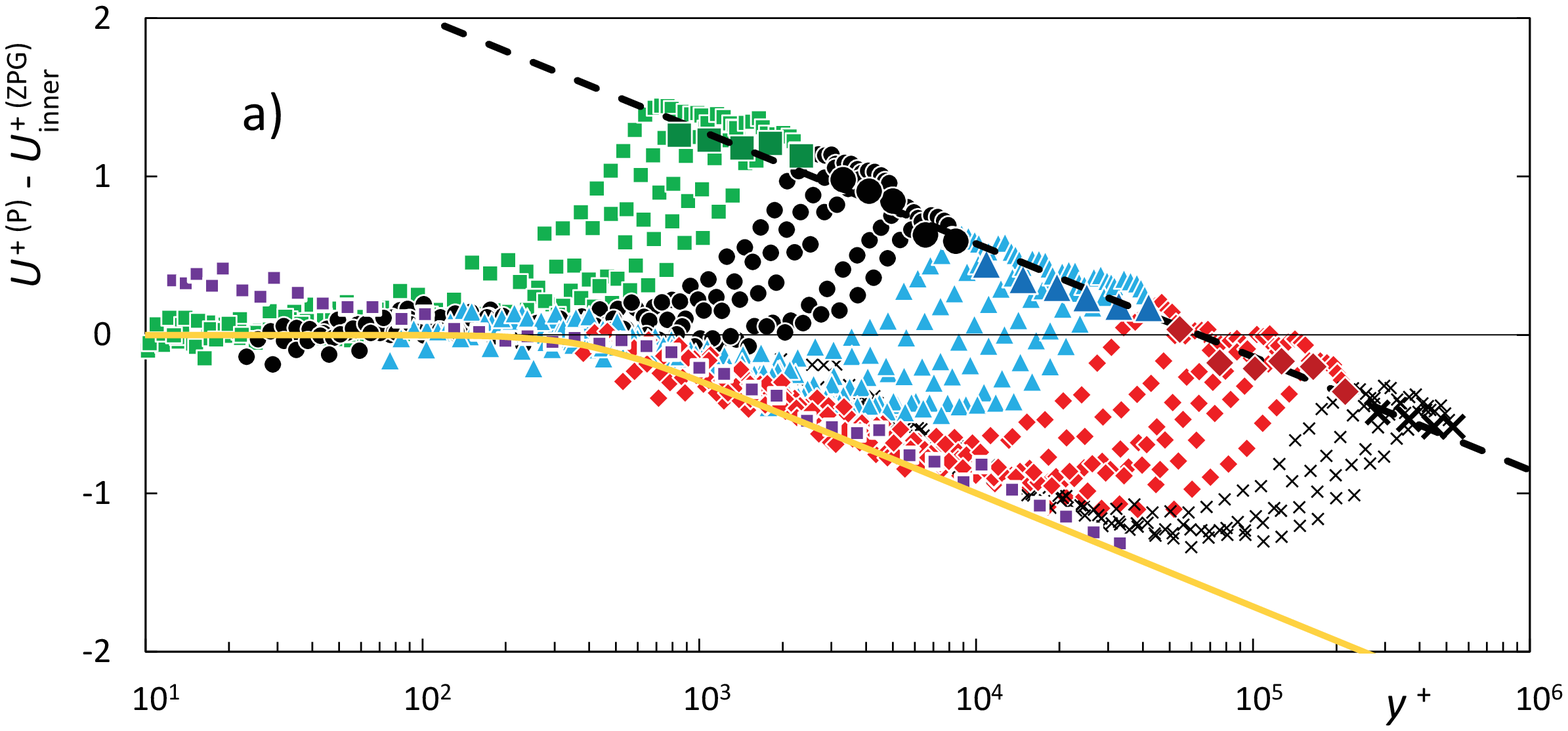}
\includegraphics[width=0.8\textwidth]{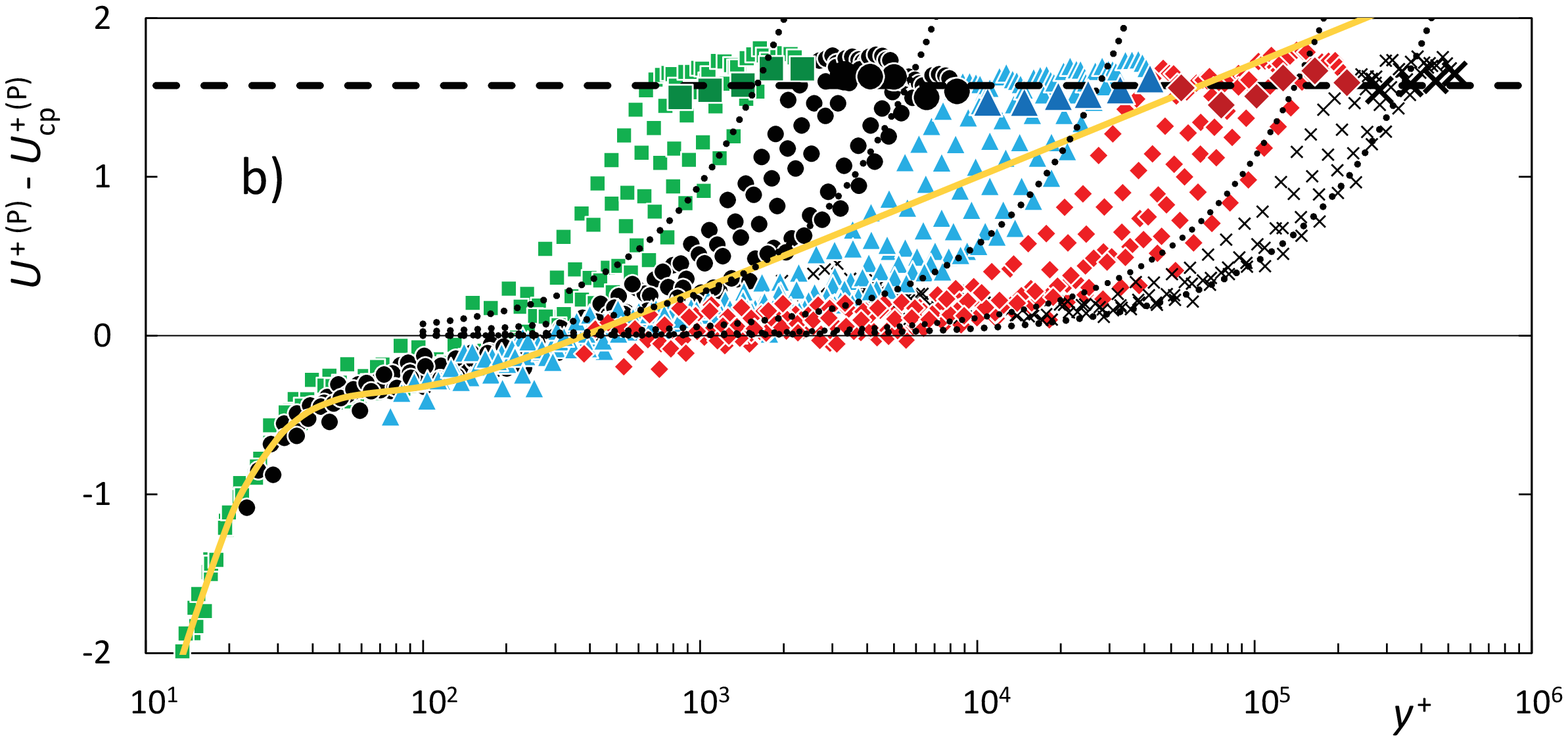}
\caption{(color online) Analysis of the overlap layer of the 26 Superpipe profiles of \cite{ZS98}, corrected according to equations (\ref{Zagcorr}) and (\ref{Hama}), for $851 \leqslant R^+ \leqslant 528000$. \textcolor{green}{$\blacksquare$}, $R^+ < 3\times 10^3$ ; $\bullet$, $3\times 10^3 < R^+ = < 10^4$ ; \textcolor{blue}{$\blacktriangle$}, $10^4 < R^+ < 5\times 10^4$ ; \textcolor{red}{$\blacklozenge$}, $5\times 10^4 < R^+ < 2.5\times 10^5$ ; $\times$, $2.5\times 10^5 < R^+$ where roughness effects become significant. Corresponding large symbols mark the centerline fitted by equ. (\ref{CLlog}) (\textbf{- - -}).
\newline (a) \textcolor{green}{$\blacksquare$} $\bullet$ \textcolor{blue}{$\blacktriangle$} \textcolor{red}{$\blacklozenge$} $\times$, $(U^{+\,\mathrm{(Z\&S)}} - U^{+\,\mathrm{(ZPG)}}_{\mathrm{inner}})$ [equ. (\ref{Musker})]. \textcolor{violet}{$\blacksquare$}, $U^{+\,\mathrm{(Z\&S)}}$, taken from fig. 17 of \cite{ZS98},  minus $U^{+\,\mathrm{(ZPG)}}_{\mathrm{inner}}$. $- - -$, $(U^{+\,\mathrm{(P)}}_{\mathrm{CL}} - U^{+\,\mathrm{(ZPG)}}_{\mathrm{inner}})$ ; \textcolor{orange}{\textbf{---}}, $(U^{+\,\mathrm{(P)}}_{\mathrm{inner}} - U^{+\,\mathrm{(ZPG)}}_{\mathrm{inner}})$ [equ. (\ref{Pinner})].
\newline (b) \textcolor{green}{$\blacksquare$} $\bullet$ \textcolor{blue}{$\blacktriangle$} \textcolor{red}{$\blacklozenge$} $\times$, $(U^{+\,\mathrm{(Z\&S)}} - U^{+\,\mathrm{(P)}}_{\mathrm{cp}})$ [equ. (\ref{Pcp})]. $- - -$, $(U^{+\,\mathrm{(P)}}_{\mathrm{CL}} - U^{+\,\mathrm{(P)}}_{\mathrm{cp}}) = 1.56$ ; \textcolor{orange}{\textbf{---}}, $(U^{+\,\mathrm{(ZPG)}}_{\mathrm{inner}} - U^{+\,\mathrm{(P)}}_{\mathrm{cp}})$ ;  $\cdot \cdot \cdot$, departure $L^{+\,\mathrm{(P)}}$ (equ. \ref{Lpart}) from the log-law for the last profile in each group.}
\label{Fig:Plog1}
\end{figure}

The different asymptotic regions for the data of \cite{ZS98}, corrected according to equations (\ref{Hama}) and (\ref{Zagcorr}), are shown in figure \ref{Fig:Plog1}. In panel (a), only the ZPG TBL inner profile (\ref{Musker}) is subtracted and the data are seen to switch around $y^+ \approx 400$ from the ZPG log-law with $\kappa_0=0.384$ to the ``true'' leading order overlap log-law with the pipe-specific $\kappa=0.436$
\begin{equation}
U^{+\,\mathrm{(P)}}_{\mathrm{cp}} = \frac{1}{0.436}\, \ln(y^+) + 6.07
\label{Pcp}
\end{equation}
which is the common part ``cp'' of the inner and outer expansions. Note that relative to  \cite{Monk17}, the switch from the ZPG log-law to the log-law (\ref{Pcp}) has been lowered from 500 to 400. As an aside, the original corrected data in fig. 17 of \cite{ZS98} are seen in figure \ref{Fig:Plog1}a to be under-corrected below $y^+ \approx 100$ .

The complete leading order inner expansion for the pipe is now modelled as in \cite{Monk17} by
\begin{eqnarray}
\label{Pinner}
U^{+\,\mathrm{(P)}}_{\mathrm{inner}} &=& U^{+\,\mathrm{(ZPG)}}_{\mathrm{inner}} + \frac{1}{3}\left(\frac{1}{0.436} - \frac{1}{0.384}\right) \ln\left[1 + (0.0025 \,y^+)^3 \right] \\ \nonumber
&\to &  U^{+\,\mathrm{(P)}}_{\mathrm{cp}} \quad \mbox{for} \quad \left(0.0025 \, y^+\right) \gg 1 \quad ,
\end{eqnarray}
with $U^{+\,\mathrm{(P)}}_{\mathrm{cp}}$ given by equ. (\ref{Pcp}). As dictated by asymptotic matching principles, the overlap log-law (\ref{Pcp}) has the same $\kappa=0.436$ as the centerline log-law
\begin{equation}
U^{+\,\mathrm{(P)}}_{\mathrm{CL}} = \frac{1}{0.436}\, \ln(R^+) + 7.63 \quad .
\label{CLlog}
\end{equation}
This is again evident in figure \ref{Fig:Plog1}b, where only the common part $U^{+\,\mathrm{(P)}}_{\mathrm{cp}}$ (equ. \ref{Pcp}) has been subtracted from the data to show the excellent collapse of all the data below $y^+ \approx 400$ onto the Musker-Chauhan profile (\ref{Musker}), as intended with the correction scheme of section \ref{sec:corrPitot}.

\begin{figure}
\center
\includegraphics[width=0.5\textwidth]{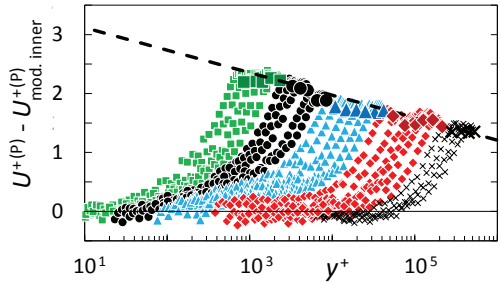}
\caption{(color online) Same data as in fig. \ref{Fig:Plog1} minus $U^{+\,\mathrm{(Z\&S)}}_{\mathrm{inner}}$ with a breakpoint between the remnant of the ZPG log-law and the pipe overlap log-law modified from $(0.0025\,y^+_{\mathrm{break}}) = 1$ to $0.7\,y^+_{\mathrm{break}}/(R^+)^{1/2} = 1$ in equation (\ref{Pinner}).
 $- - -$, resulting centerline $\kappa_{\mathrm{CL}} = 0.47$. }
\label{Fig:Plog1alt}
\end{figure}

\textcolor{black}{Finally, in view of the different proposals for the start of the pipe overlap log-law, for instance $y^+_{\mathrm{break}}/(R^+)^{1/2} = 3$ proposed by \cite{MMHS13} and others, the present scaling $(0.0025\,y^+_{\mathrm{break}}) = 1$ for the start of the overlap log-law in equation (\ref{Pinner}) and figure \ref{Fig:Plog1}a requires a closer examination. Replacing $[0.0025\,y^+]$ in the logarithm of equation (\ref{Pinner}) by $[0.7\,y^+(R^+)^{-0.5}]$, for instance, modifies figure \ref{Fig:Plog1}b to figure \ref{Fig:Plog1alt}, where $\kappa_{\mathrm{CL}}=0.47$ is no longer equal to the overlap $\kappa$ of 0.436, as required by the leading order matching between the overlap and outer profiles. The reason for this is easily identified by evaluating the limit of the modified equation (\ref{Pinner}) for $(0.7\,y^+/\sqrt{R^+}) \gg 1\,$, resulting in a modification of the common part (equ. \ref{Pcp}) to $(1/0.436)\ln(y^+) + 4.31 + 0.16\,\ln(R^+)$. Subtracting this modified common part from the data yields a wake function which decreases as $0.16\,\ln(R^+)$ on the centerline, corresponding to a centerline kappa of $[(1/0.436) - 0.16]\ln(R^+) = 0.47\,\ln(R^+)$, as seen in figure \ref{Fig:Plog1alt}. This argument against a Reynolds number dependent $y^+_{\mathrm{break}}$, i.e. the lower end of the overlap log-law, is independent of the actual value of the overlap kappa and the power $p\neq 0$ of $R^+$ in $[y^+_{\mathrm{break}} (R^+)^{-p}] =$ constant. In practice, the problem ``disappears'' in the data uncertainty for sufficiently small $p$, but must remain of concern when speculating about the infinite Reynolds number limit.}

\subsection{\label{sec:lin}The linear part of the outer ``wake'' profile}

\begin{figure}
\center
\includegraphics[width=0.65\textwidth]{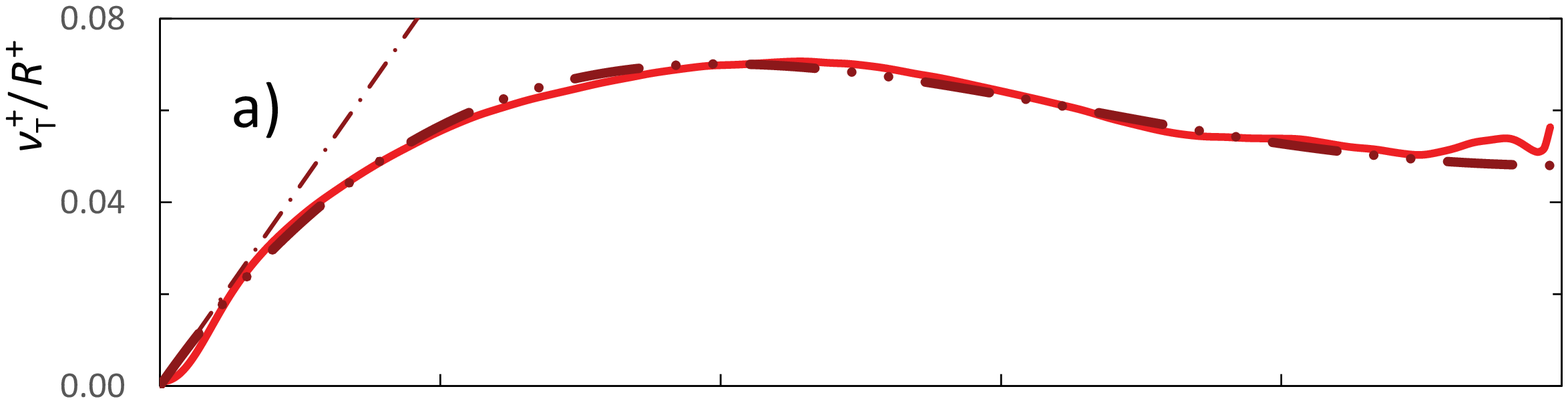}
\includegraphics[width=0.65\textwidth]{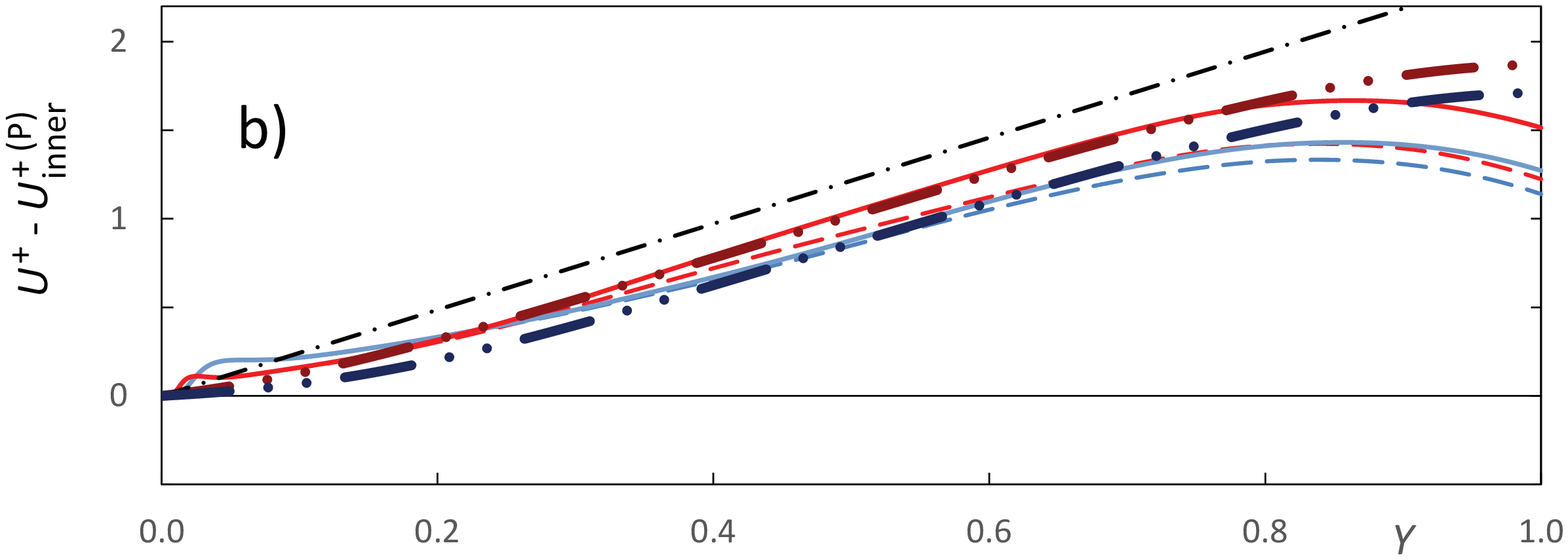}
\caption{(color online) (a) \textcolor{red}{---}, turbulent viscosity $N_T \equiv \mu_T^+/R^+$, calculated from the DNS of \cite{ElKhoury2013} for $R^+=999$; \textcolor{red}{$-\cdot -$}, $0.436 Y$; \textcolor{red}{$-\cdot \cdot -$}, fit (\ref{NTfit}). (b) DNS mean velocity profiles of \cite{ElKhoury2013} for $R^+=550$ (\textcolor{blue}{---}) and $R^+=999$ (\textcolor{red}{---}) minus $U^{+\,\mathrm{(P)}}_{\mathrm{inner}}$ (equ. \ref{Pinner}); - - -, corresponding profiles minus $U^{+\,\mathrm{(ZPG)}}_{\mathrm{inner}}$ (equ. \ref{Musker}); $- \cdot \cdot -$, $L^{+\,\mathrm{(P)}}$ fitted by equation (\ref{Lpart}) for $R^+ = 550$ and 999; $- \cdot -$, limit $2.43 Y$ at $R^+=\infty$ (without the slope decrease towards the CL).}
\label{Fig:NT}
\end{figure}

Driven by the desire to obtain more reliable values for the K\'arm\'an parameter from data at moderate Reynolds numbers, \cite{Yajnik70}, \cite{Afzal76,AfzalIUTAM96}, \cite{Jimenez07} and \cite{Luchini17}, among others, have proposed various higher order corrections to the log-law, notably linear corrections $\propto y^+/R^+$. While most of these proposed corrections were justified by asymptotic matching arguments, it is argued here that the linear correction in the overlap layer is just the tail of the linear part of the outer wake profile, and therefore dependent on the outer boundary conditions. This is demonstrated with the pipe DNS of \cite{ElKhoury2013}. It is straightforward to determine the effective turbulent viscosity from the momentum equation and the
computed mean velocity derivative. Beyond the inner region it is, in outer variables, equal to $N_T \equiv \nu^+_T/R^+ = (1-Y)(\dd U^+/\dd Y)^{-1}$ \citep[see e.g.][section 3.5]{Wilcox}. As shown in figure \ref{Fig:NT}a, $N_T$ is well fitted by
\begin{equation}
N_T = 0.048 + 0.122 (1-Y)^2 - 0.17 (1-Y)^4 \cong 0.436 Y - 0.898 Y^2 + \mathcal{O}(Y^3) ~~\mbox{for}~~ Y\ll 1~~.
\label{NTfit}
\end{equation}

For small $Y$, the fit (\ref{NTfit}) corresponds to
\begin{equation}
U^+ = \ln{Y}/0.436 + C + 2.43 Y + \mathcal{O}(Y^2) ~~\mbox{for}~~ Y\ll 1~~.
\label{Ufit}
\end{equation}
From figure \ref{Fig:NT}b it is obvious that the linear term in equation (\ref{Ufit}) needs to be offset by an estimated $(80/R^+)$ to obtain good fits at the lower $R^+$. This is achieved by the construction
\begin{eqnarray}
\label{Lpart}
L^{+\,\mathrm{(P)}} &=& \frac{1}{c_s}\,\ln\left[1 - \mbox{e}^{- 2.43\, c_s Y_0} + \mbox{e}^{2.43\, c_s (Y-Y_0)}\right] - 2.43\, F_3(Y)~~\mbox{with} \\
c_s = 5, Y_0 &=& 80/R^+ \ \mbox{and} \  F_n(Y) \equiv \frac{1}{n \ln^{n-1}\left(\frac{\pi}{2}\right)}\,\left\{\ln[\frac{\pi}{2} Y] - \ln\left[\sin\left(\frac{\pi}{2} Y\right)\right]\right\}^n  ,
\label{defF}
\end{eqnarray}
where $c_s < \infty$ smoothes the corner at $Y=Y_0$ and the constants in the argument of the logarithm in equation (\ref{Lpart}) ensure that $L^{+\,\mathrm{(P)}}(0) = 0$. For large $R^+$, the first logarithmic term of $L^{+\,\mathrm{(P)}}$ quickly approaches the straight line $2.43Y$, shown in figure \ref{Fig:NT}b. The zero centerline slope $(\dd L^{+\,\mathrm{(P)}}/\dd Y)(1)=0$ is obtained with the function $F_n(Y)$ which has a slope of unity on the centerline and is proportional to $Y^{2n}$ for $Y\ll 1$. The choice of $F_3$ in equation (\ref{Lpart}) ensures that, after the offset by $(80/R^+)$, $L^{+\,\mathrm{(P)}}$ remains linear up to $Y\approx 0.5$.

It is noted in passing that the two wake profiles in figure \ref{Fig:NT}b, obtained from the DNS of \cite{ElKhoury2013}, show some suspicious ``humps'' near the origin. They are clearly the result of an imperfect ``overshoot'' term in the Musker-Chauhan fit, the last term of equation (\ref{Musker}), which should go to zero faster below $y^+=30$ and be higher at these low $R^+$.  However, since the appropriate higher order corrections in the inner region, presumably of $\mathcal{O}(1/R^+)$ like the quadratic term $-(y^+)^2/(2 R^+)$ in the Taylor expansion of $U^+$ about the origin, have no direct bearing on the present analysis, no effort has been made here to improve equation (\ref{Musker}).

The present construction of the linear term in the outer expansion shows that its coefficient depends on the entire outer shape of the turbulent viscosity $N_T$ and not on some matching condition in the overlap layer. For pipe and channel, $N_T$ must be symmetric about the centerline. Assuming that it can be represented as a polynomial in powers of $(1-Y)^2$, as in equation (\ref{NTfit}), a linear term in the outer velocity profile is unavoidable, except in the unlikely case that the coefficients of all the $Y^2$-terms of $N_T$ sum up to $-\kappa$. The situation is different for the ZPG TBL where the departure from the log-law is $\propto Y^4$, according to the data analysis of \cite{Monk17}, implying a small-$Y$ expansion of $N_T$ of the form $\kappa_0 Y + \mathcal{O}(Y^5)$.

The discussion on how the slope of 2.43 for the linear part of the pipe wake might be connected to the pressure gradient parameter $\beta = - \widehat{R}\, \widehat{p_x}/ \widehat{\tau_{\mathrm{wall}}}$ is postponed to the concluding section \ref{sec:conclusion}.

\subsection{\label{sec:outer}The complete outer pipe profile}

\begin{figure}
\center
\includegraphics[width=0.65\textwidth]{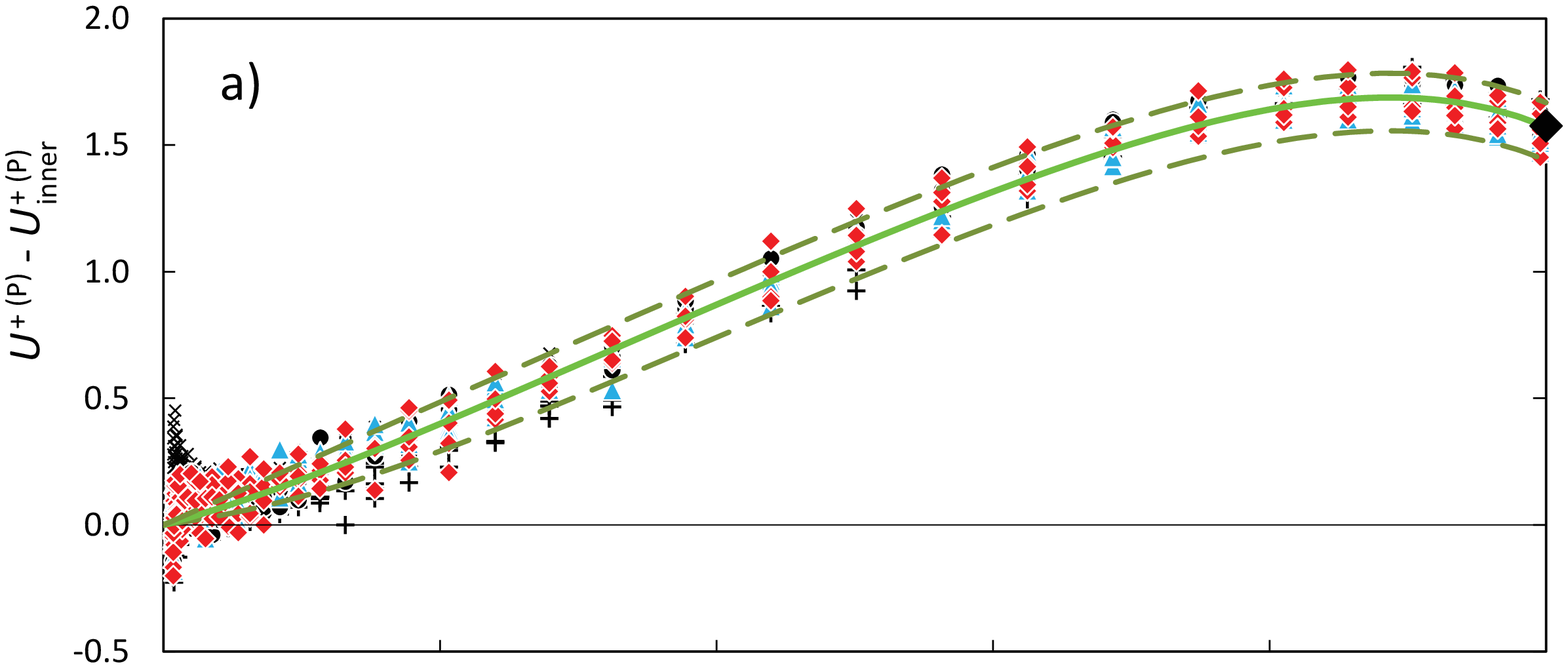}
\includegraphics[width=0.65\textwidth]{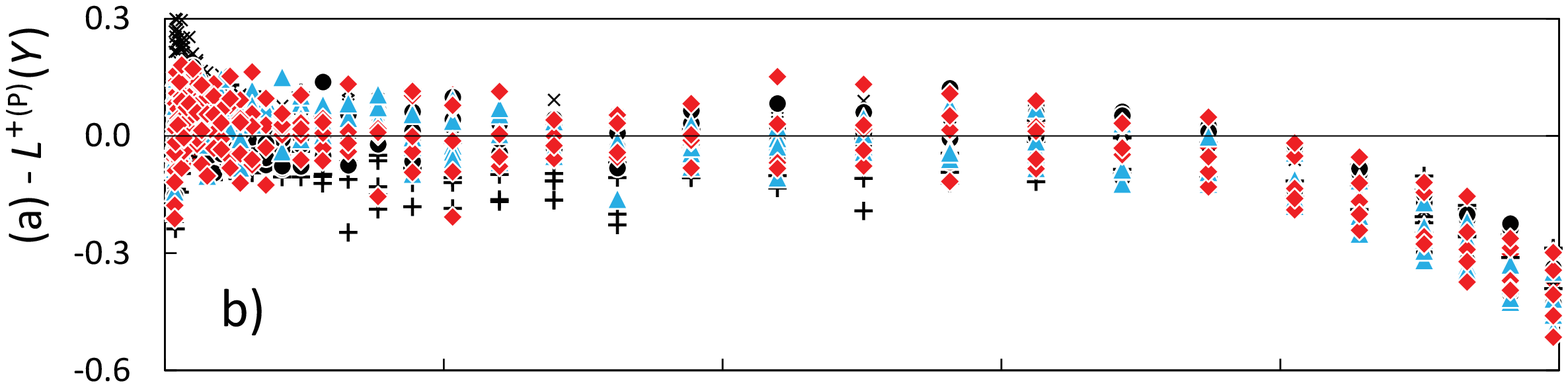}
\includegraphics[width=0.65\textwidth]{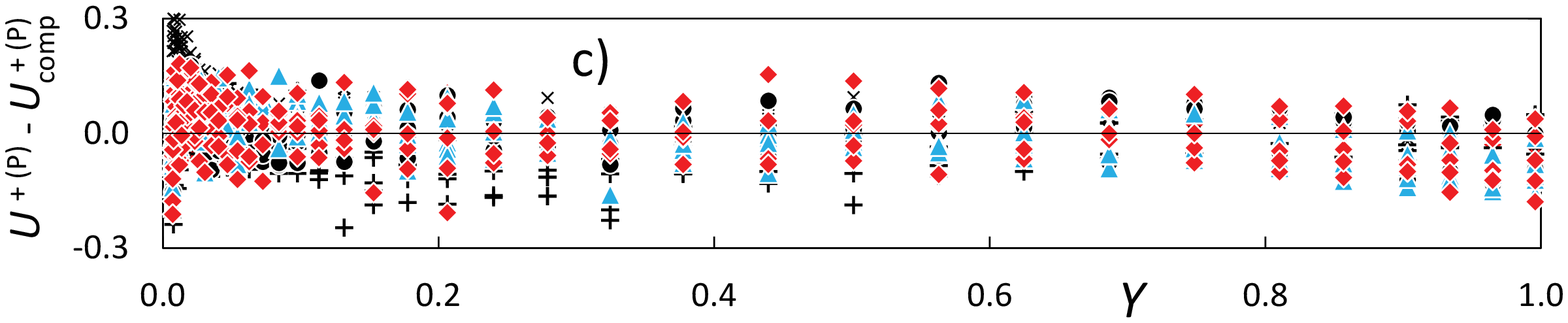}
\caption{(color online) (a) The wake function $U^{+\,\mathrm{(P)}}_{\mathrm{wake}}(Y)$ of equation (\ref{Pcomp2}) with the same data and color coding as in figure \ref{Fig:Plog1}; \textcolor{green}{- - -, ---, - - -}, asymptotically linear part $L^{+\,\mathrm{(P)}}$ in equation (\ref{Lpart}) for $R^+$ = 850, 2000 and $\infty$; $\blacklozenge$ , $U^{+\,\mathrm{(P)}}_{\mathrm{CL}}$ [equ. \ref{CLlog}] minus $U^{+\,\mathrm{(P)}}_{\mathrm{inner}}(y^+ = R^+)$. (b) $U^{+\,\mathrm{(P)}}_{\mathrm{wake}}(Y) - L^{+\,\mathrm{(P)}}(Y;R^+=2000)$. (c) Complete fit $U^{+\,\mathrm{(Z\&S)}} - U^{+\,\mathrm{(P)}}_{\mathrm{comp}}$ [equs. (\ref{Pcomp}), (\ref{Pcomp2})].}
\label{Fig:Plog2}
\end{figure}

The obvious question is now whether the term $L^{+\,\mathrm{(P)}}$ of equation (\ref{Lpart}), deduced from the DNS of \cite{ElKhoury2013}, also ``works'' for the Z\&S data. Figure \ref{Fig:Plog2}a shows that it does, and that the Z\&S data all fall between the $L^{+\,\mathrm{(P)}}$ for the lowest $R^+$ of 850 and $R^+=\infty$. Since the data scatter in figure \ref{Fig:Plog2}a is of the same magnitude as the difference between these two $L^{+\,\mathrm{(P)}}$, only an average $L^{+\,\mathrm{(P)}}$ corresponding to $R^+=2000$ is subtracted from the data to obtain figure \ref{Fig:Plog2}b which shows that the pipe wake is well described by $L^{+\,\mathrm{(P)}}$, i.e. is linear in the interval $2Y_0 \lessapprox Y\lessapprox 0.5\,$.

The last step towards the composite expansion is to compensate the centerline slope of $- (1/\kappa)$ caused by the log-law in $U^{+\,\mathrm{(P)}}_{\mathrm{inner}}$. This is easily achieved by adding $(1/0.436) F_3(Y)$ to the data of figure \ref{Fig:Plog2}b , \textcolor{black}{resulting}
 in the rather satisfactory fit of  $U^{+\,\mathrm{(P)}}_{\mathrm{comp}}$ \textcolor{black}{shown} in figure \ref{Fig:Plog2}c, \textcolor{black}{and} given by
\begin{eqnarray}
\label{Pcomp}
U^{+\,\mathrm{(P)}}_{\mathrm{comp}} &=& U^{+\,\mathrm{(P)}}_{\mathrm{inner}} + U^{+\,\mathrm{(P)}}_{\mathrm{outer}} - U^{+\,\mathrm{(P)}}_{\mathrm{cp}} \quad \mbox{with} \\
\textcolor{black}{U^{+\,\mathrm{(P)}}_{\mathrm{wake}} \equiv} U^{+\,\mathrm{(P)}}_{\mathrm{outer}} - U^{+\,\mathrm{(P)}}_{\mathrm{cp}} &=& L^{+\,\mathrm{(P)}}(Y; R^+=2000) + (1/0.436)\,F_3(Y) ~,
\label{Pcomp2}
\end{eqnarray}
with $F_3$ defined by equation (\ref{defF}). \textcolor{black}{It is worth reiterating here that the leading behavior of the present wake fit (\ref{Pcomp2}) is \textbf{linear} in $Y$, with a higher order offset. This is qualitatively different from previous fits, including the one in \cite{Monk17}, which were all variations of Coles' sin$^2$ wake function, i.e. $\propto Y^2$ for small $Y$. Only at low $R^+$ there is some resemblance between the new and the traditional pipe wake functions due to the offset in equation (\ref{Lpart}). }

\section{\label{sec:conclusion}Conclusions}

In conclusion, the extreme simplicity of the present Pitot correction of the original Superpipe data of \cite{ZS98} strongly supports the notion of \cite{Monk17} that the near-wall velocity profile in the pipe is, within experimental uncertainty and up to small $\mathcal{O}(1/R+)$ corrections, equal to the ZPG TBL profile out to $y^+_{\mathrm{break}} \approx 400-500$, where the pipe profile switches to the overlap log-law (\ref{Pcp}) with a pipe-specific K\'arm\'an parameter $\kappa$ of 0.436. Comparing with the Superpipe data of \cite{MLJMS04} used in \cite{Monk17}, the difference between overlap $\kappa$'s - 0.436 versus 0.421 - is on the high side of uncertainty estimates \citep[see e.g.][]{Bailey14}. One reason may be that the data set of \cite{ZS98} contains more low Reynolds number profiles than the one of \cite{MLJMS04}. Eliminating the Z\&S profiles below $R^+ = 10^4$ does reduce their centerline $\kappa_{\mathrm{CL}}$ from 0.436 to 0.430, but in view of the increasing uncertainty of the Pitot corrections with Reynolds number and the centerline $\kappa_{\mathrm{CL}}$ of 0.446 recently found in the CICLoPE pipe by \cite{TSFP17}, the original overlap $\kappa$ of 0.436 appears to be a reasonable value for the pipe and has been maintained for the present analysis.

The important finding here is that the sequence of asymptotic regions is the \textbf{same} for the two Superpipe data sets and the structure of the fitting functions for the different asymptotic regions is identical \textcolor{black}{to the one in \cite{Monk17}, despite the different $\kappa$'s. The substantial difference to previous fits is the new wake function which is asymptotically linear over about half the pipe radius and represents a radical departure from the traditional $\sin^2$ type function, originally introduced by \cite{Coles56}. This asymptotically linear outer departure from the log-law suggests that the linear higher order corrections to the log-law \citep[see e.g.][and references therein]{Luchini17} are slaved to the outer expansion and not an intrinsic consequence of asymptotic matching.}

It is now natural to ask if and how the slope of $L^{+\,\mathrm{(P)}}$ is related to the pressure-gradient parameter $\beta = - \widehat{R}\, \widehat{p_x}/ \widehat{\tau_{\mathrm{wall}}}$. Since the wake of ZPG TBL's shows no linear part, it is tempting to make the slope of $L^{+\,\mathrm{(P)}}$ proportional to $\beta$, at least for $\beta \leq \mathcal{O}(1)$. However, the geometry and in particular transverse wall curvature are known to \textcolor{black}{also} modify the asymptotic structure and are therefore expected to contribute to the slope of $L^{+\,\mathrm{(P)}}$ in a non-trivial way. In particular flat plate TBL's in weak pressure gradients are expected to have a linear \textcolor{black}{wake} part different from pipe and channel, because of the different free stream boundary conditions for the turbulent viscosity $N_T$.

\begin{figure}
\center
\includegraphics[width=0.65\textwidth]{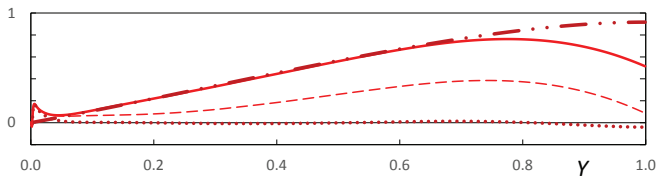}
\caption{(color online) \textcolor{red}{---}, DNS mean velocity profile $U^{+\,\mathrm{(L\&M)}}$ of \cite{LM14} for $H^+=5200$ minus $U^{+\,\mathrm{(Ch)}}_{\mathrm{inner}}$ ; \textcolor{red}{- - -}, DNS profile minus $U^{+\,\mathrm{(ZPG)}}_{\mathrm{inner}}$ (equ. \ref{Musker}); \textcolor{red}{$- \cdot \cdot -$}, $L^{+\,\mathrm{(Ch)}}$ fitted by equation (\ref{LpartCh}); \textcolor{red}{$\cdot \cdot \cdot$}, $U^{+\,\mathrm{(L\&M)}} - U^{+\,\mathrm{(Ch)}}_{\mathrm{comp}}$. For an explanation of the nonphysical ``hump'' near the origin, see the comment in section \ref{sec:lin} regarding the same phenomenon in figure \ref{Fig:NT}b.}
\label{Fig:Plog4}
\end{figure}

To close the discussion, the channel DNS of \cite{LM14} for $H^+ = 5200$ is reanalyzed with the methodology of the present paper. Maintaining the overlap $\kappa$ at 0.413 as in \cite{Monk17}, but shifting the break between the ZPG and channel log-laws to $y^+=500$ yields the wake profile of figure \ref{Fig:Plog4}. As for the pipe, it is first fitted by $L^{+\,\mathrm{(Ch)}}$, given by
\begin{equation}
\label{LpartCh}
L^{+\,\mathrm{(Ch)}} = \frac{1}{c_s}\,\ln\left[1 - \mbox{e}^{- 1.23\, c_s Y_0} + \mbox{e}^{1.23\, c_s (Y-Y_0)}\right] - 1.23\, F_2(Y)~\mbox{with}~
c_s = 5, Y_0 = 150/H^+
\end{equation}
with $F_2$ defined by equation (\ref{defF}). Adding just $(1/0.413) F_3(Y)$ to fix the centerline slope leads to a near perfect composite fit. It is intriguing to find that the slopes of the linear wake component in pipe and channel, fitted here as 2.43 and 1.23, are related by a factor of 2. It is left to the reader to speculate whether the two slopes should be fitted by $1.22 \beta$ or by a more complicated function of $\beta$, transverse wall curvature, $\kappa(\beta)$ or other parameters.

\begin{acknowledgments}
The author is most grateful to Hassan Nagib for his helpful comments on various points addressed in this paper.
\end{acknowledgments}

\bibliographystyle{jfm}
\bibliography{Karmanaps}

\end{document}